\pdfoutput=1
 \documentclass[aps,prd,reprint,twocolumn,showpacs,superscriptaddress]{revtex4-1}  

\usepackage{tabularx}
\makeatletter
\def\hlinewd#1{%
\noalign{\ifnum0=`}\fi\hrule \@height #1 %
\futurelet\reserved@a\@xhline}
\makeatother
 \usepackage{auncial}
\usepackage{scrextend}
\usepackage{relsize}
\usepackage{amsmath}
\usepackage{amssymb}
\usepackage{epsfig}
\usepackage{graphicx}
\usepackage{hyperref}
\usepackage{dcolumn}   % needed for some tables
\usepackage{tabu}
\usepackage{boldline}
\usepackage{slashed}
\usepackage{multirow}
\usepackage{bbold}
\usepackage{color}
\usepackage[normal]{subfigure}
\usepackage{rotating}
\usepackage[margin=0.9in,a4paper]{geometry}
\usepackage[table]{xcolor}
\usepackage{enumitem}
\usepackage{colortbl}
\usepackage{array,multirow}
\usepackage{cellspace}
\definecolor{nicered}{rgb}{0.7,0.1,0.1}
\definecolor{nicegreen}{rgb}{0.1,0.5,0.1}
\definecolor{red}{rgb}{1.0, 0, 0}

\definecolor{Grn}{rgb}{0.,0.75,0.}
\definecolor{Blu}{rgb}{0.,0.,1.}

\def\arg{\mbox{arg}\,}

\def\gsim{\raise0.3ex\hbox{$\;>$\kern-0.75em\raise-1.1ex\hbox{$\sim\;$}}}
\def\lsim{\raise0.3ex\hbox{$\;<$\kern-0.75em\raise-1.1ex\hbox{$\sim\;$}}}
\def\mb[#1]{\mathbf{#1}}
\definecolor{LightCyan}{rgb}{0.88,1,1}
\definecolor{piggypink}{rgb}{0.99, 0.87, 0.9}
\definecolor{applegreen}{rgb}{0.55, 0.71, 0.0}
\definecolor{darkpastelgreen}{rgb}{0.01, 0.75, 0.24}
\definecolor{green-yellow}{rgb}{0.68, 1.0, 0.18}

\newcommand{\meV}{{\, \rm meV}}

\newcommand{\be}{\begin{equation}}
\newcommand{\ee}{\end{equation}}
\newcommand{\bea}{\begin{eqnarray}}
\newcommand{\eea}{\end{eqnarray}}
\begin{document}
% ----------------- preprint numbers ------------------
%\begin{frontmatter}

%%%%%%%%%%%%%%%%%%%%%% 
%\Red{{\bf Only in the arXiv version:}
\begin{flushright}
{\footnotesize IFIC/20-44, FTUV-20-0921.9442, P3H-20-047, TTP20-032}
\end{flushright}

% ------------- Title and authors ---------------------

\title{Neutrino Observables from a U(2) Flavor Symmetry}

%\title{Neutrino Predictions of a realistic U(2) Model of Flavor}

\author{Matthias Linster}
%\email{}
\affiliation{\normalsize \it 
Institut f\"ur Theoretische Teilchenphysik, Karlsruhe Institute of Technology, Karlsruhe, Germany
}

\author{Jacobo Lopez-Pavon}
%\email{}
\affiliation{\normalsize \it 
Instituto  de  F\'isica  Corpuscular,  Universidad de Valencia  and  CSIC,Edificio  Institutos  Investigaci\'on,  Catedr\'atico  Jos\'e  Beltr\'an  2,  46980,  Spain}

\author{Robert Ziegler}
%\email{robert.ziegler@cern.ch}
\affiliation{\normalsize \it 
Institut f\"ur Theoretische Teilchenphysik, Karlsruhe Institute of Technology, Karlsruhe, Germany
}

% ------------------------------------------------------
\begin{abstract}
\noindent
We study the predictions for CP phases and absolute neutrino mass scale for broad classes of models with a U(2) flavor symmetry. For this purpose we consider the same special textures in neutrino and charged lepton mass matrices that are succesful in the quark sector. While in the neutrino sector the U(2) structure enforces two texture zeros, the contribution of the charged lepton sector to the PMNS matrix can be parametrized by two rotation angles. Restricting to the cases where at least one of these angles is small, we obtain three representative scenarios. In all scenarios we obtain a narrow prediction for the sum of neutrino masses in the range  of 60$-$75 meV, possibly in the reach of upcoming galaxy survey experiments. All scenarios can be excluded if near-future experimental date provide evidence for either neutrinoless double-beta decay or inverted neutrino mass ordering.
\end{abstract}

%%%%%%%%%%  \pacs{14.80.Va, 14.65.Jk}
%%%%%%%%%%  \keywords{}
% \end{keyword}
%%  \PACS 
%\end{frontmatter}

\maketitle

%%%%%%%%%%%%%%%%%%%%%% 
\section{Introduction} 
Neutrino oscillation experiments have firmly established that neutrinos have tiny masses and large mixings. We currently know with good level of precision the values of the three mixing angles parametrizing the PMNS lepton mixing matrix $U(\theta_{12}$, $\theta_{13}$ and $\theta_{23}$), and the absolute value of the squared neutrino mass differences $\Delta m^2_{ij}\equiv m_i^2-m_j^2$. There are just four  questions left open: i) Is CP violated in the lepton sector? ii) Is the neutrino mass ordering normal (NO) or inverted (IO)? iii) What is the absolute neutrino mass scale? iv) Are neutrinos Dirac or Majorana particles?

Information about the CP violating phase and the mass ordering, albeit not yet statistically conclusive, is starting to be extracted from neutrino experiments. Global analyses~\cite{Esteban:2020cvm, nuFIT} show a mild preference for NO, and a $\sim3\sigma$ hint for CP violation from the T2K~\cite{T2K:2020} and NOvA~\cite{NOvA:2020} long baseline neutrino oscillation experiments, if the analysis is restricted to IO. For NO, there is a very slight tension between T2K and NOvA data with best fits at $\delta\approx250^\circ$ and $\delta\approx150^\circ$ respectively. Near future experiments such as T2HK and DUNE are expected to provide a definite answer for the neutrino mass ordering and a measurement of the CP phase $\delta$ with an uncertainty of $~10^\circ-20^\circ$ at $1\sigma$.

On the other hand the strongest constraints on the overall neutrino mass scale currently arise from cosmology. Specifically, the analysis of the Cosmic Microwave Background (CMB) and Baryon Acoustic Oscillations (BAO) data gives $\sum m_i <0.12\,\rm{eV}$ at $95\%\,\rm{CL}$~\cite{Aghanim:2018eyx}. Future galaxy surveys such as Euclid~\cite{Amendola:2016saw} and DESI~\cite{Aghamousa:2016zmz} will have a sensitivity $\sigma\left(\sum m_i\right)\approx 0.02$ eV  and are expected to report a measurement in the next $\sim 5-10$ years. Even better sensitivity could be achieved with the next generation of ground-based CMB experiments~\cite{Abazajian:2019eic}.

The best Laboratory bound is given by KATRIN~\cite{Aghanim:2018eyx} which has recently reported the upper limit $m_\beta\equiv \sqrt{\sum_i|U_{ei}|^2m_i^2} <1.1$ eV ($90\%$ CL) and is expected to probe the region $m_\beta \gtrsim 0.2$ eV ($90\%$ CL) in the near future, while the Project 8 Neutrino-Mass Experiment has a target future sensitivity of $0.04$ eV~\cite{Project8:2020}.

If neutrinos are Majorana particles, also neutrinoless double beta ($0\nu\beta\beta$) decay experiments are sensitive to the absolute neutrino mass scale (and Majorana phases of the PMNS matrix) through the ``effective Majorana mass" $m_{\beta\beta}\equiv\sum_i U_{ei}^2 m_i$. The most stringent present bound was reported by KamLAND-Zen and is given by the range $m_{\beta\beta}<0.061-0.165$ eV ($90\%$  CL)~\cite{KamLAND-Zen:2016pfg}, depending on the input considered for the nuclear matrix elements.  The goal of next generation $0\nu\beta\beta$ decay experiments like LEGEND~\cite{Abgrall:2017syy} and nEXO~\cite{Albert:2017hjq} is to probe the entire IO region ($m_{\beta\beta} \gtrsim 10^{-2}$ eV).

In this work, we make the attempt to predict the undetermined neutrino observables using a U(2) flavor symmetry (for similar approaches see Ref.~\cite{Feruglio:2019ktm} and references therein). This symmetry provides a particularly simple and elegant framework to express all hierarchies in fermion masses and mixings in terms of just two small parameters (the two spurions responsible for breaking the U(2) flavor symmetry), apart from various ${\cal O}(1)$ coefficients.  The U(2) structure typically leads to a characteristic pattern of fermion mass matrices $m_{ij}$, with $m_{11} = m_{13} = m_{31} = 0$ and $m_{21} = - m_{12}$, which makes this framework predictive despite the presence of the additional ${\cal O}(1)$ parameters. This feature will allow us to study the consequences for neutrino observables in a broad class of U(2) models, without the need of specifying  explicit flavor quantum numbers or Lagrangians.

The first U(2) models have been proposed in Refs.~\cite{Barbieri:1995uv, Barbieri:1997tu} in the context of supersymmetry (SUSY) and flavor quantum numbers compatible with a unified $SO(10)$ gauge group. In this case the special structure of quark mass matrices leads to the exact prediction $V_{ub}/V_{cb} = \sqrt{m_u/m_c}$, which is strongly disfavored by data. Indeed the original $SO(10)$  models  with a U(2) flavor symmetry (along with a D$_3$$\times$U(1) variant of it~\cite{Dermisek:1999vy}) were essentially ruled out with the advent of the B-factories~\cite{Roberts:2001zy}. However, it is still possible to construct viable U(2) models with flavor quantum numbers compatible only with a unified $SU(5)$ gauge group, which gives an excellent fit to the quark and charged lepton sector~\cite{DGPZ, Linster:2018avp}. These models also do not require SUSY, i.e. they are viable with a single light Higgs doublet in contrast to the original $SO(10)$ models.  They can be also employed to predict the flavor structure of new light physics like $Z^\prime$~\cite{FNZ} gauge bosons or vector leptoquarks~\cite{Barbieri:2019zdz}, which allow to address the persisting anomalies in semi-leptonic B-decays.

Neutrinos can be easily incorporated in the $SU(5)$-compatible U(2) framework~\cite{Linster:2018avp}, either as Dirac or Majorana neutrinos. For Dirac neutrinos right-handed neutrinos with suitable charges can be introduced, which gives a  viable neutrino mass matrix with the same pattern as quarks and leptons, but anarchic (i.e.~all entries of the same order). The smallness of neutrino masses is here a result of a strong overall suppression by flavor  breaking spurions. A more predictive possibility is to consider instead of U(2) = SU(2)$\times$U(1) the flavor group D$_6$ $\times$U(1), where the (discrete) dihedral group D$_6$ essentially mimickes the $SU(2)$ group structure~\footnote{Taking a discrete subgroup of SU(2) is also suggested to remove the associated massless Goldstone bosons. Instead the Goldstone boson of the U(1) factor is welcome as it plays the rule of the QCD axion~\cite{Ema:2016ops, Calibbi:2016hwq, Linster:2018avp}.}, but with \emph{symmetric} doublet contraction, which implies the very same structure of fermion mass matrices as in the U(2) case, but with $m_{21} = + m_{12}$. The Weinberg operator then induces Majorana neutrino masses with a parametric flavor suppression that is \emph{predicted} by the charged lepton sector, and naturally leads to an anarchic structure, which allows for an excellent fit to all fermion masses and mixings including the neutrino sector~\cite{Linster:2018avp}. 

In this work we want to study this remarkable feature in more detail. For this purpose we will not construct an explicit flavor model and fit parameters as in Ref.~\cite{Linster:2018avp}, but rely only on the particular textures~\footnote{In non-SUSY models these textures are only approximate, however the corrections are typically tiny, of order $V_{cb}^2$~\cite{Linster:2018avp}.}  of neutrino and charged lepton sector motivated by the flavor symmetry~\footnote{Although this structure follows from D$_6$ $\times$ U(1) rather from SU(2) $\times$ U(1), we refer to it as the ``U(2)'' texture for simplicity, because for quark and charged lepton masses and mixings the different sign does not play any role.}, i.e. $m_{11} = m_{13} = m_{31} = 0$ and $m_{21} =  m_{12}$. In this way we can pin down the relevant implications for neutrino observables for a broad class of U(2) models, without the need of specifying (and defining)  ``${\cal O}(1)$" parameters. 

In the neutrino sector this approach reproduces the well-studied case of two texture zeros, also called the ``$A_2$" texture~\cite{Frampton:2002yf, Guo:2002ei, Dev:2006qe, Fritzsch:2011qv, Meloni:2012sx, Kitabayashi:2015jdj, Zhou:2015qua, Singh:2016qcf, Alcaide:2018vni, Xing:2019vks}. In the limit when there is no contribution from mixing in the charged lepton sector, this structure leads to a phenomenologically viable PMNS matrix with  predictions for Dirac and Majorana phases and the overall mass scale, which are only limited by the uncertainty of the input parameters. 

In general the contribution of the charged lepton sector spoils the predictivity of neutrino textures. However, in the present scenario  the imposed U(2) textures  reduce the number of free rotation angles from 6 to 2, which can be chosen to be the left-handed (LH) and right-handed (RH) rotations in the 2-3 sector. We will restrict here to the two limiting cases where one of  these rotation angles is small, which is motivated by explicit models where these rotations are related to small CKM angles~\cite{DGPZ, Linster:2018avp, Barbieri:2019zdz, Calibbi:2020jvd}. As we will show, these models lead to one-parameter deviations from the viable $A_2$ texture, and are thus still predictive.

%%%%%%%%%%%%%%%%%%%%%% 
\section{Setup}

\subsection{Neutrino Sector} 
In the neutrino sector we consider a general Majorana  mass matrix with a U(2) texture, given by
\begin{align}
m_{\nu} =  \begin{pmatrix}0 & m^\nu_{12} &0  \\%
						   m^\nu_{12} & m^\nu_{22}  & m^\nu_{23}  \\%
						  0 & m^\nu_{23} & m^\nu_{33} \end{pmatrix} = V_\nu^*  {\rm diag} (m_1, m_2, m_3)   V_\nu^\dagger \, .
 \label{def}
\end{align}
Compared to a generic Majorana mass matrix there are two conditions, which lead to relations among neutrino masses and mixing angles. The neutrino mixing matrix $V_\nu$ can be parametrized in the standard PDG form 
\begin{align}
V_\nu &= P_\nu^\prime V_{23}^\nu V_{13}^\nu V_{12}^\nu P_\nu \, , 
\label{parametrization}
\end{align}
with phase matrices  $P_\nu = {\rm diag} (e^{i \alpha^\nu_1}, e^{i \alpha_2^\nu}, 1), P^\prime_\nu = {\rm diag} (e^{i \beta^\nu_1}, e^{i \beta^\nu_2}, e^{i \beta^\nu_3}) $ and  unitary matrices $V_{ij}^\nu$ describing rotations in the $i-j$ plane parametrized by angles in the first quadrant $s_{ij}^\nu \equiv \sin \theta_{ij}^\nu \ge0, c_{ij}^\nu \equiv \cos \theta_{ij}^\nu \ge0$, with the Dirac phase $\delta^\nu$ contained in $V_{13}^\nu$
\begin{gather}
V^\nu_{23}  = \begin{pmatrix} 1 &  0 &  0 \\ 0 &  c_{23}^\nu & s_{23}^\nu  \\  0 & - s_{23}^\nu & c_{23}^\nu \end{pmatrix} \, , \qquad V^\nu_{12}  = \begin{pmatrix} c^\nu _{12}  & s_{12}^\nu   & 0   \\   -s_{12}^\nu   & c_{12}^\nu  & 0 \\  0 & 0 & 1 \end{pmatrix}\, , \nonumber \\ 
V^\nu_{13}   = \begin{pmatrix}  c_{13}^\nu  & 0 & s_{13}^\nu  \, e^{- i \delta^\nu  }\\  0 & 1 & 0 \\ - s_{13}^\nu  \,  e^{i \delta^\nu  } & 0 & c_{13}^\nu  \end{pmatrix} \, .
\end{gather}
From Equation~\eqref{def}, one finds from the conditions $(m_\nu)_{11} = (m_\nu)_{13} = 0$  the two complex equations
\begin{gather}
\label{eq1}
 \frac{m_1}{m_3} e^{ i \tilde{\alpha}_1} +(t_{12}^{\nu})^2 \frac{m_2}{m_3} e^{ i \tilde{\alpha}_2}  + \frac{(t_{13}^{\nu})^2}{ (c_{12}^{\nu})^2}  e^{ - i \delta^\nu} = 0 \, ,  \\
 \frac{m_1}{m_3} e^{ i \tilde{\alpha}_1}  - \frac{m_2}{m_3}  e^{ i \tilde{\alpha}_2}  + \frac{t_{13}^{\nu} }{c_{13}^{\nu} c_{12}^{\nu} s_{12}^{\nu} t_{23}^{\nu}}  = 0 \, , 
\label{eq2}
\end{gather}
with $t^\nu_{ij} \equiv \tan \theta_{ij}^\nu$ and $\tilde{\alpha}_i\equiv 2\alpha_i^\nu+\delta^\nu$. These equations can be easily solved analytically: the real and imaginary part of Eq.~\eqref{eq2} give two equations that can be solved for $\cos\tilde{\alpha}_i$ as a function of neutrino masses $m_i$, mass differences $\Delta m^2_{ij} \equiv m_i^2 - m_j^2$ and mixing angles $\theta_{ij}^\nu$
\bea
\label{eq:alpha1}
\cos\tilde{\alpha}_1&=&\frac{\Delta m^2_{21} - m_3^2 A^2 }{2 m_1 m_3 A} \, , 
\\
\cos\tilde{\alpha}_2 &=&\frac{\Delta m^2_{21} + m_3^2 A^2 }{2  m_2 m_3 A } \, , 
\label{eq:alpha2}
\eea
with the shorthand
\begin{align}
A = \frac{t_{13}^{\nu}  }{c_{13}^{\nu} c_{12}^{\nu} s_{12}^{\nu} t_{23}^{\nu}} \, .
\end{align}
From the real and imaginary part of Eq.~\eqref{eq1} one can  obtain a solution for $\cos \delta^\nu$ and an equation without phases
\begin{align}
\label{eq:delta}
\cos \delta^\nu &= \frac{A}{2 (t_{13}^{\nu})^2}  \left[ 1  - 2 (s_{12}^{\nu})^2  - \frac{\Delta m^2_{21}}{m_3^2 A^2} \right] \, , \\
(t_{13}^{\nu})^4 &= \frac{m_1^2}{m_3^2} + (s_{12}^{\nu})^2 \frac{\Delta m^2_{21}}{m_3^2} - (s_{12}^{\nu})^2  (c_{12}^{\nu})^2 A^2 \, .
\label{eq:m}
\end{align}
It is not guaranteed that a solution always exists, but if it does, it is given by Eqs.~\eqref{eq:alpha1}, \eqref{eq:alpha2}, \eqref{eq:delta} and \eqref{eq:m}. Finally, one can show that $\sin \tilde{\alpha}_i$ must have the same sign as $\sin \delta^\nu$, which fixes the signs of $\tilde{\alpha}_i$ in terms of the sign of $\delta^\nu$, which is undetermined. 

By eliminating $A$ from from Eqs.~\eqref{eq:delta} and \eqref{eq:m}, one can  furthermore derive an equation that gives the allowed range for $m_3$ (and therefore a range for the absolute mass scale):
\begin{gather}
\label{range}
m_3^{\rm min} \le m_3 \le  m_3^{\rm max} \, , \\
m_3^{\rm max, min } \equiv \frac{(c_{12}^\nu)^2 }{(t_{13}^\nu)^2} \left| m_1 \pm m_2 (t_{12}^\nu)^2 \right| \, .
\end{gather}  
It is instructive to consider the special case where the charged lepton sector does not contribute to the PMNS matrix, so that 
$U= V_\nu$, and the physical mixing angles and phases are given by the neutrino sector quantities above, $\theta_{ij} = \theta_{ij}^\nu$ and $\delta = \delta^\nu$, $\alpha_i = \alpha_i^\nu$. We will refer to this case as the {\bf ``Diagonal Charged Lepton" (DCL) Scenario} in the following. 

In this scenario the measured values of the mixing angles and mass differences directly determine four neutrino observables, as a result of the two complex constraints on the neutrino mass matrix: the overall neutrino mass scale in Eq.~\eqref{eq:m}, the Dirac phase from Eq.~\eqref{eq:delta}, and the Majorana phases from Eqs.~\eqref{eq:alpha1} and \eqref{eq:alpha2}. Moreover,  Eq.~\eqref{def} states that the effective Majorana mass vanishes, $m_{\beta \beta} = |\sum_i m_i U_{ei}^2|= |(m_{\nu})_{11}| =0$, which directly implies (see Fig.~\ref{loglog}) that an inverted neutrino mass hierarchy is not viable and the sum of neutrino masses \emph{automatically} satisfies the stringent bounds from cosmology. Indeed, taking the central values for mixing angles from the recent global fit in Ref.~\cite{Esteban:2020cvm} (including SK atmospheric data), there is no solution for the inverted hierarchy case, as can be seen by expanding Eq.~\eqref{eq:m} to leading order in $s_{13}^2 \approx 0.02$, giving $s_{13}/t_{23} > m_1/m_3$, which cannot be satisfied if $m_3 < m_1$. 

Instead for normal hierarchy a solution exists (see also Table~\ref{table} and Fig.~\ref{CPs23DCL}), which predicts for the Dirac phase the two possible values $\delta \approx  (77^\circ, 283^\circ)$, which fall inside the $1\sigma$ intervals~\cite{Esteban:2020cvm} recenty reported by NOvA~\cite{NOvA:2020} and T2K~\cite{T2K:2020} respectively. The total sum of neutrino masses is fixed to be $\Sigma  \approx 65 \meV$, in agreement with the present cosmological bound $\sum m_i <0.12\,\rm{eV}$ at $95\%\,\rm{CL}$~\cite{Aghanim:2018eyx} and in the reach of near future galaxy surveys as Euclid~\cite{Amendola:2016saw} and DESI~\cite{Aghamousa:2016zmz}. The effective neutrino mass is predicted to be $m_\beta \approx 10  \meV $,  too small to be measured in the near future by KATRIN or even by Project 8. These predictions serve as useful reference values, because they are obtained whenever the charged lepton sector gives only small corrections to the PMNS matrix. 

\subsection{Charged Lepton Sector} 
Also the charged lepton mass matrix is taken as a general matrix with a U(2) texture, 
\begin{align}
M_{e} =  \begin{pmatrix}0 & m^e_{12} &0  \\%
						   m^e_{12} & m^e_{22}  & m^e_{23}  \\%
						  0 & m^e_{32} & m^e_{33} \end{pmatrix} \, .
						  \label{Me}
\end{align}
This matrix is diagonalized by bi-unitary rotations $V_{eL}^\dagger M_e V_{eR} = M_e^{\rm diag}$, where the matrix $V_{eL}$ can be parametrized analogously to the neutrino sector as in Eq.~\eqref{parametrization}, with $\theta^\nu_{ij} \to \theta^{Le}_{ij}, \delta^\nu \to \delta^{Le}, \alpha^\nu_i \to \alpha^{Le}_i, \beta^\nu_i \to \beta^{Le}_i$, and similar for the matrix $V_{eR}$. 
Because of the special structure of the mass matrix in Eq.~\eqref{Me}, these parameters are not independent. One can show~\cite{FNZ} that the 3+3 rotation angles in the LH and RH rotations depend only on two real parameters $\theta^{Le}_{23}$ and $\theta^{Re}_{23}$, which  correspond to the 2-3 rotation angles in the LH and RH sectors. Similar constraints hold for the phases, so that the 
number of mixing parameters is reduced by eight, indeed corresponding to four complex constraints on the mass matrix. 

A further simplification can be made because in motivated scenarios at least one of the mixing angles  $\theta^{Le}_{23}$ or $\theta^{Re}_{23}$ can be small. For example in unified scenarios both parameters are directly related to the corresponding mixing angles in the down quark sector, $\theta^{Ld}_{23}$ and $\theta^{Rd}_{23}$, which are essentially fixed in order to reproduce quark masses and mixings. The left-handed rotation angle is CKM-like,  $\theta^{Ld}_{23} \sim V_{cb} \sim 0.04$, while the right-handed rotation angle has to be large,  $\theta^{Rd}_{23} \sim 1$. In the following, we therefore consider the two simple cases in which either $\theta^{Le}_{23}$ or $\theta^{Re}_{23}$ is small $(\lesssim V_{cb})$ (if both are small we recover the case with diagonal charged leptons). The first scenario, which can be realized in U(2) models with fermions unified within a Pati-Salam gauge group~\cite{Barbieri:2019zdz}, the ``U(2)$_{\rm PS}$" scenario, and the second scenario, which can be realized in models with fermions unified within a $SU(5)$ gauge group~\cite{Linster:2018avp}, the ``U(2)$_5$" scenario. In both models, the structure of charged lepton rotations simplifies drastically, and only depends on a single rotation angle. Neglecting small angles ${\cal O}(V_{cb})$, one has in the two scenarios:
\begin{itemize}
\item{ \bf ``$\mathbf{U(2)_{{\bf PS}}}$" Scenario:}  ($s^{Re}_{23}$ free, $s^{Le}_{23} \ll 1$)
\begin{align}
s_{13}^{Le} &  \ll 1 \, , & s_{12}^{Le} =  \sqrt{\frac{c_{23}^{Re} m_e}{ m_\mu}} \, ,
\end{align} 
valid as long as $c_{23}^{Re} \gtrsim m_e/m_\mu$. Up to phases,  the charged lepton sector rotation in the ${\rm U(2)_{PS}}$ scenario is given by
\begin{align}
V_{eL} = V^e_{12} = \begin{pmatrix} c^{Le}  _{12}  & s_{12}^{Le}   & 0   \\   -s_{12}^{Le}    & c_{12}^{Le}   & 0 \\  0 & 0 & 1 \end{pmatrix} \, .
\label{U2PS}
\end{align}
\item { \bf ``$\mathbf{U(2)_{5}}$" Scenario:}  ($s^{Le}_{23}$ free,  $s^{Re}_{23} \ll 1$)
\begin{align}
s_{13}^{Le} &  \ll 1 \, , &  s_{12}^{Le} =  \sqrt{\frac{m_e}{c_{23}^{Le} m_\mu}} \, ,
\end{align} 
valid as long as $c_{23}^{Le} \gtrsim m_e/m_\mu$. 
Up to phases,  the charged lepton sector rotation in the ${\rm U(2)_{5}}$ scenario is given by
\begin{align}
V_{eL} = V^e_{23} V^e_{12} = \begin{pmatrix} c^{Le}  _{12}  & s_{12}^{Le}   & 0   \\   - c_{23}^{Le}s_{12}^{Le}    & c_{23}^{Le} c_{12}^{Le}   & s_{23}^{Le} \\  s_{23}^{Le} s_{12}^{Le} & - s_{23}^{Le} c_{12}^{Le} & c_{23}^{Le} \end{pmatrix} \, .
\label{U25}
\end{align}
\end{itemize}
In both scenarios also the number of free phases is reduced when considering only the physical parameters entering the PMNS matrix, as discussed in the next section. There are two new phases in the ${\rm U(2)_{5}}$ scenario, corresponding to two rotations in the 2-3 and 1-2 plane, while there is one new phase in the ${\rm U(2)_{PS}}$ scenario, associated with a single rotation in the 1-2 plane. Note that this rotation angle is bounded from above in the  ${\rm U(2)_{PS}}$ scenario, $s_{12}^{Le} \le \sqrt{m_e/m_\mu} \approx 0.07$, while it can be maximal in the ${\rm U(2)_{5}}$ scenario.

\subsection{PMNS Matrix}
The PMNS matrix is given by $
U= V_{eL}^\dagger V_\nu$, and can be decomposed as in Eq.~\eqref{parametrization}, with  $\theta^\nu_{ij} \to \theta_{ij}, \delta^\nu \to \delta, \alpha^\nu_i \to \alpha_i, \beta^\nu_i \to \beta_i$. As $V_{eL}$ is unique only up to arbitrary shifts of the phases $\alpha^{Le}_i \to \alpha^{Le}_i + \phi_i$ with $\phi_i \in [ 0 , 2 \pi]$, and $V_\nu$ is unique only up to sign shifts of the Majorana phases $\alpha^\nu_i \to \alpha^\nu_i + S_i$ with $S_i = \pm \pi$,  the PMNS matrix is unique only up to sign and phase shifts $\beta_{1,2} \to \beta_{1,2} + \phi_i$ and $\alpha_i \to \alpha_i + S_i$. This residual freedom can be used to set $\beta_i= 0$ and consider only $\alpha_i \in [0, \pi]$. 

For a given scenario for the charged lepton sector, Eq.~\eqref{U2PS} or \eqref{U25}, one can then relate the neutrino sector parameters $\theta^\nu_{ij}, \delta^\nu, \alpha^\nu_i, \beta^\nu_i$ to the physical (and partially measured) parameters $\theta_{ij}, \delta, \alpha_i$, up to a few parameters describing the charged lepton sector, see Appendix~\ref{appU2PS} and ~\ref{appU25}. For the ${\rm U(2)_{PS}}$ scenario there is one free angle $s_{23}^{Re}$ and one free effective phase $\beta$, while for the ${\rm U(2)_{5}}$ scenario there is one free angle $s_{23}^{Le}$ and two free effective phases $\beta_{1}, \beta_{2}$, cf. also Table~\ref{table}. For given values of these free parameters, together with the input on physical mixing angles and  mass differences from neutrino oscillation data, one can use equations Eqs.\eqref{eq:alpha1}-\eqref{eq:m} to calculate the physical Majorana and Dirac phases along with the overall neutrino mass scale. This in turn allows to calculate the sum of neutrinos masses $\sum m_i$, the effective Majorana mass $m_{\beta \beta}$, and the effective neutrino mass $m_\beta$. As we will discuss in the next section, the resulting predictions fall in a quite narrow window, as a result of the rather constraining nature of the underlying U(2) flavor symmetry, even if not fully specified.
%%%%%%%%%%%%%%%%%%%%%% 
\section{Results} 
In order to analyze the predictions in the different scenarios we proceed as follows: {\bf 1)} for a given mass ordering (NO or IO) we use the NuFIT distributions provided in Ref.~\cite{Esteban:2020cvm} to draw a random sample of $\theta_{13}, \theta_{23}, \theta_{12}, \Delta m_{21}^2$ and $\Delta m_{31}^2$ (NO) or $\Delta m_{32}^2$ (IO); {\bf 2)} for   scenario U(2)$_{\rm PS}$ [U(2)$_{5}$] we  draw a random sample of $\theta^{Re}_{23}, \beta$ [$\theta^{Le}_{23}, \beta_1, \beta_2$] assuming flat distributions; {\bf 3)} we use Eqs.~\eqref{A1}-\eqref{A3} [Eqs.~\eqref{B1}-\eqref{B3}] in order to obtain the resulting values of neutrino sector angles $s_{13}^\nu, s_{23}^\nu, s_{12}^\nu$ and phases $\alpha_i^\nu$ and $\delta^\nu$, as a function of the PMNS phases $\alpha_i$ and $\delta$; {\bf 4)}  we use Eqs.~\eqref{eq:alpha1}-\eqref{eq:m} in order to solve for the Majorana phases $\alpha_i$, the Dirac phase $\delta$ and the lightest neutrino mass (a solution might not exist), and derive the observables $m_\beta, m_{\beta \beta}$ and $\sum m_i$. For the DCL scenario we skip steps {\bf 2)} and {\bf 3)}, since there are no free parameters in the charged lepton sector and PMNS angles and phases coincide with neutrino sector angles and phases. 

Note that we are treating the CP phase $\delta$ always as an output, i.e.~we ignore that partial information about this parameter is provided by the global fits to oscillation experiments. However, since its error is still quite large, and moreover there is slight tension between T2K and NOvA, we think that this procedure is appropriate for the moment. When the uncertainty on $\delta$ substantially decreases in the future this parameter should also be treated  as an input. 

Our results in the two scenarios and the DCL reference point (i.e. the pure $A_2$ texture with no contribution to the PMNS matrix from the charged lepton sector) are summarized in Table~\ref{table}, where we display the most likely values of $\sum m_i, m_\beta, m_{\beta \beta}$ along with their $1\sigma$, $2\sigma$ and $3\sigma$ ranges, defined by containing 68.3\%, 95.4\% and 99.7\% of the generated sample points, respectively. 

\begin{table*}[ht]
\centering
\setlength{\extrarowheight}{2pt}
\begin{ruledtabular}
\begin{tabular}{cccccc}
Scenario & Free parameters & NO/IO &  $\sum m_i$~[meV] & $m_\beta$~[meV]  & $m_{\beta \beta}$~[meV]  \\
\colrule
DCL & none & NO & $65.0_{-0.6}^{+0.9}$ & $10.0_{-0.2}^{+0.3}$ & $0_{-0}^{+0}$ \\
%& & & $(64.4 \rightarrow 65.9)_{1 \sigma}$ & $(9.8 \rightarrow 10.3)_{1 \sigma}$ & $(0 \rightarrow 0)_{1 \sigma}$ \\
%& & & $(63.7 \to 67.5)_{2 \sigma}$ & $(9.6 \to 10.9)_{2 \sigma}$ & $(0 \to 0)_{2 \sigma}$\\
& & & $(64 \to 68)_{2 \sigma}$ & $(10 \to 11)_{2 \sigma}$ & $(0 \to 0)_{2 \sigma}$\\
& & & $(63 \to 69)_{3 \sigma}$ & $(9 \to 12)_{3 \sigma}$ & $(0 \to 0)_{3 \sigma}$\\
\colrule
${\rm U(2)_{PS}}$ &  $s_{23}^{Re}, \beta$ & NO & $65.7_{-2.1}^{+3.8}$ & $9.8_{-0.3}^{+1.6}$ & $1.2_{-0.3}^{+0.5}$ \\
& & & $(62 \to 72)_{2 \sigma}$ & $(9 \to 13)_{2 \sigma}$ & $(0 \to 2)_{2 \sigma}$\\
& & & $(62 \to 75)_{3 \sigma}$ & $(9 \to 14)_{3 \sigma}$ & $(0 \to 2)_{2 \sigma}$\\
\colrule
${\rm U(2)_{5}}$ & $s_{23}^{Le}, \beta_{1}, \beta_2$ & NO & $63.7_{-2.1}^{+4.4}$ & $9.5_{-0.3}^{+1.5}$ & $1.8_{-0.8}^{+1.3}$ \\
& & & $(60 \to 74)_{2 \sigma}$ & $(9 \to 13)_{2 \sigma}$ & $(0 \to 4)_{2 \sigma}$\\
& & & $(59 \to 272)_{3 \sigma}$ & $(9 \to 85)_{3 \sigma}$ & $(0 \to 54)_{3 \sigma}$\\
\\
& & IO & $224.2_{-36.1}^{+173.8}$ & $77_{-10}^{+54}$ & $68.0_{-12.2}^{+31.0}$ \\
& & & $(173 \to 1070)_{2 \sigma}$ & $(65 \to 303)_{2 \sigma}$ & $(49 \to 255)_{2 \sigma}$\\
& & & $(167 \to 5584)_{3 \sigma}$ & $(63 \to 497)_{3 \sigma}$ & $(1 \to 299)_{3 \sigma}$\\
\end{tabular}
\end{ruledtabular}
\caption{\label{table} Predictions for neutrino observables in the three scenarios. The first line shows the most likely value along with the 1$\sigma$ interval defined to contain 68.3 \% of all generated sample points, analogously are defined the $2 \sigma$ and $3 \sigma$ regions.}
\end{table*}

\begin{figure}
\centering
\includegraphics[scale=0.26]{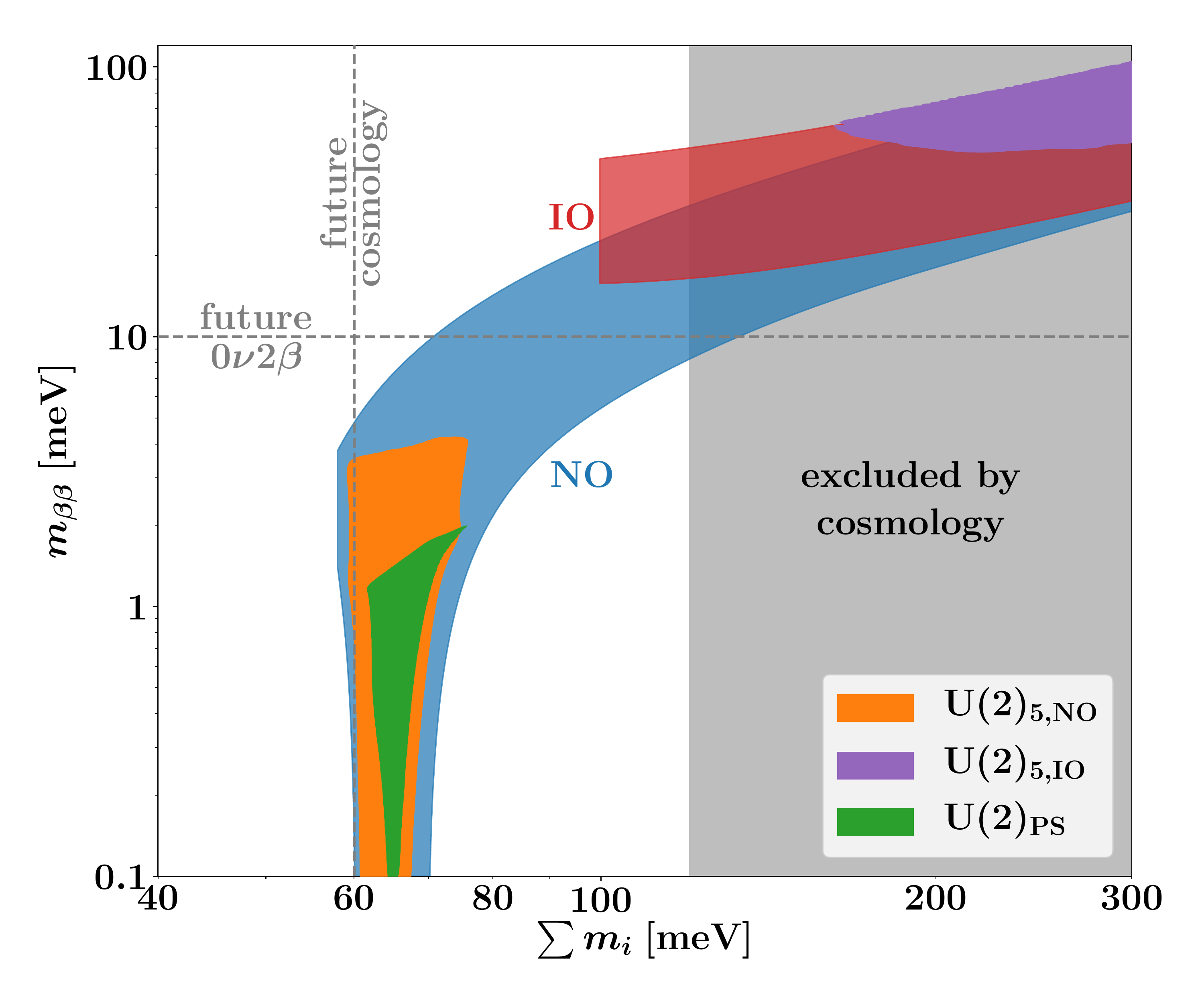}
\caption{Parameter space at $2 \sigma$ in the $m_{\beta \beta}/\sum m_i$ plane for generic NO (blue), generic IO (red), and the $\rm U(2)_{PS}$ (green), $\rm U(2)_{5, IO}$ (violet) and $\rm U(2)_{5,NO}$ (orange) scenarios. Also shown are the present constraints from cosmology~\cite{Aghanim:2018eyx} (grey region) and the expected future bounds on $m_{\beta \beta}$ and $\sum m_i$ (dashed grey lines), see text for details. \label{loglog} }  
 \end{figure}

As one can see from Table~\ref{table}, IO is only possible for scenario U(2)$_{5}$, however the entire parameter space is ruled out by the cosmological bounds~\footnote{These bounds can considerably be relaxed, but this would require neutrinos to have non-standard properties, see for instance Refs.~\cite{Lorenz:2018fzb,Chacko:2019nej, Escudero:2020ped}.} on $\sum m_i$ (see also Fig.~\ref{loglog}). Instead for NO the predictions in the two scenarios U(2)$_{5}$ and U(2)$_{\rm PS}$ are close to those of the reference point DCL, despite the presence of additional free parameters. In particular the prediction for the overall neutrino mass scale remains quite low, which implies that in both scenarios $m_{\beta \beta}$ is probably below the sensitivity of next generation $0\nu\beta\beta$ decay experiments.  Similarly, the prediction for $m_\beta$ remains very low, beyond the future reach of KATRIN. On the other hand, the low overall scale implies that at $2 \sigma$ all points  satisfy the cosmological bounds on $\sum m_i$. 

Note that the deviation from DCL is controlled by the size of the charged lepton rotation angle, which is bounded from above by $\sqrt{m_e/m_\mu}$ in the  ${\rm U(2)_{PS}}$ scenario. Therefore the  predictions in ${\rm U(2)_{PS}}$ cannot deviate much from DCL, and in particular IO is not viable. The NO parameter space in ${\rm U(2)_{PS}}$ is largely contained in the one of the  ${\rm U(2)_{5}}$ scenario, since the latter allows for larger rotation angles and thus larger deviations from DCL.  
\begin{figure}
\centering
\includegraphics[scale=0.26]{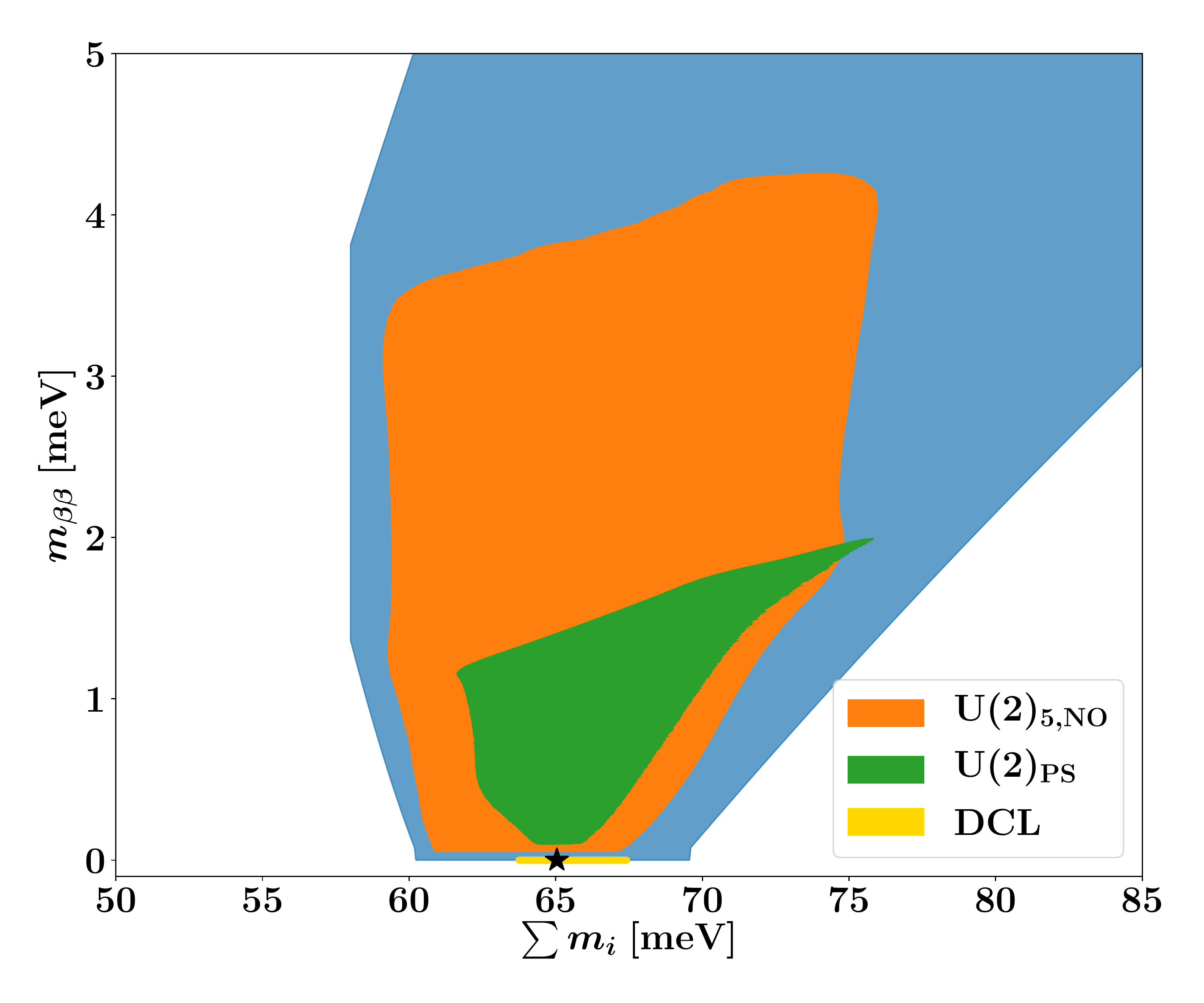}
\caption{Preferred parameter space at $2 \sigma$ in the $ m_{\beta \beta}/\sum m_i$ plane for generic NO (blue), and the diagonal charged lepton (DCL) (yellow),  $\rm U(2)_{5,NO}$ (orange) and $\rm U(2)_{PS}$ (green)  scenarios. The black star denotes the prediction in the DCL scenario for the best-fit value.  \label{linlin}}
 \end{figure}

We visualize these results in Figs.~\ref{loglog} and \ref{linlin}, which 
show the preferred parameter space (at $2 \sigma$) in the $m_{\beta \beta}/\sum m_i$ plane  for the different scenarios and the generic  (i.e. experimentally allowed) case, which has been obtained by minimizing the NuFIT likelihood at each point. While  Fig.~\ref{loglog} is the standard plot in logarithmic scale, the DCL scenario is only visible in the linear scale of Fig.~\ref{linlin} since the prediction for $m_{\beta \beta}$ identically vanishes  in the DCL case. In this plot we also denote the prediction in the DCL scenario for the best-fit value with a black cross, which shows that the deviations from this point in the two scenarios are rather small, despite the presence of the additional parameters. We do not show a similar plot for the $m_{\beta}/\Sigma m_i$ plane, since these parameters are strongly correlated and do not provide further information ($m_{\beta}$ is below future sensitivities anyway).

Finally we analyze the predictions for the CP phase $\delta$ and compare it to the present experimental situation represented by the global fit in Ref.~\cite{Esteban:2020cvm}. Since the resulting values for $\delta$ depend to a large extent on the value of $s_{23}$, we display fit and model predictions in the $\delta/s_{23}^2$ plane. 
\begin{figure}
\centering
\includegraphics[scale=0.25]{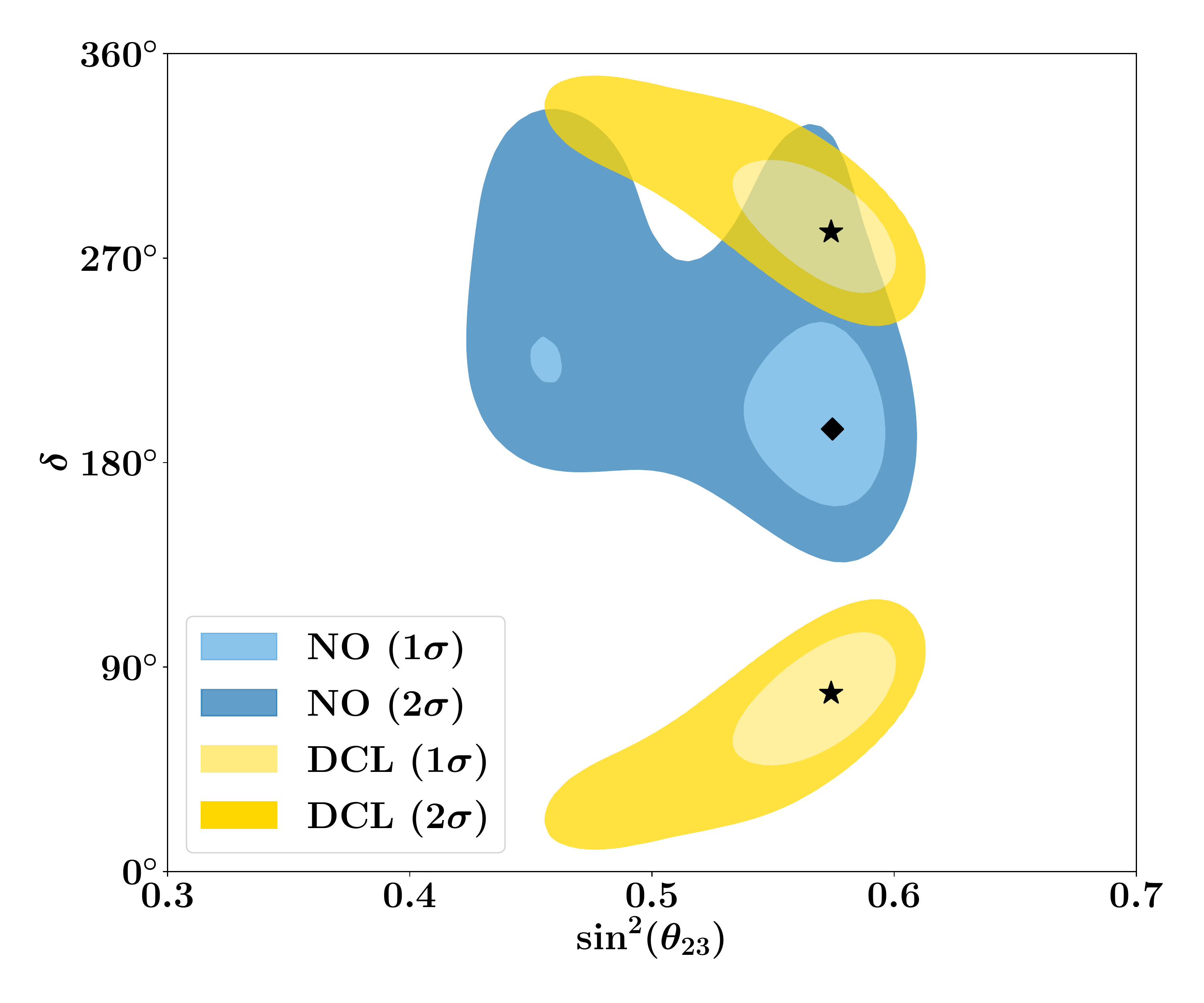}
\caption{\label{CPs23DCL}  Preferred values for the Dirac phase $\delta$ as a function of $\sin^2_{23}$  for the DCL scenario (yellow) and  the global fit in Ref.~\cite{Esteban:2020cvm} (blue). The black star denotes the most likely value in the DCL scenario, while the black diamond is the NuFIT best-fit point. }
 \end{figure}
For the DCL reference scenario this is shown in Fig.~\ref{CPs23DCL}, which highlights the predictivity of this scenario (that has no free parameters in the charged lepton sector). While at the moment there is just a slight tension (at the level of $1\sigma$) between the prediction for $\delta$ and the global fit, it will be interesting to revisit this case once there is more precise information from experiment, because future data might increase this tension and rule out this scenario (and thus the $A_2$ neutrino texture) entirely. 

\begin{figure*}
\centering
\includegraphics[scale=0.17]{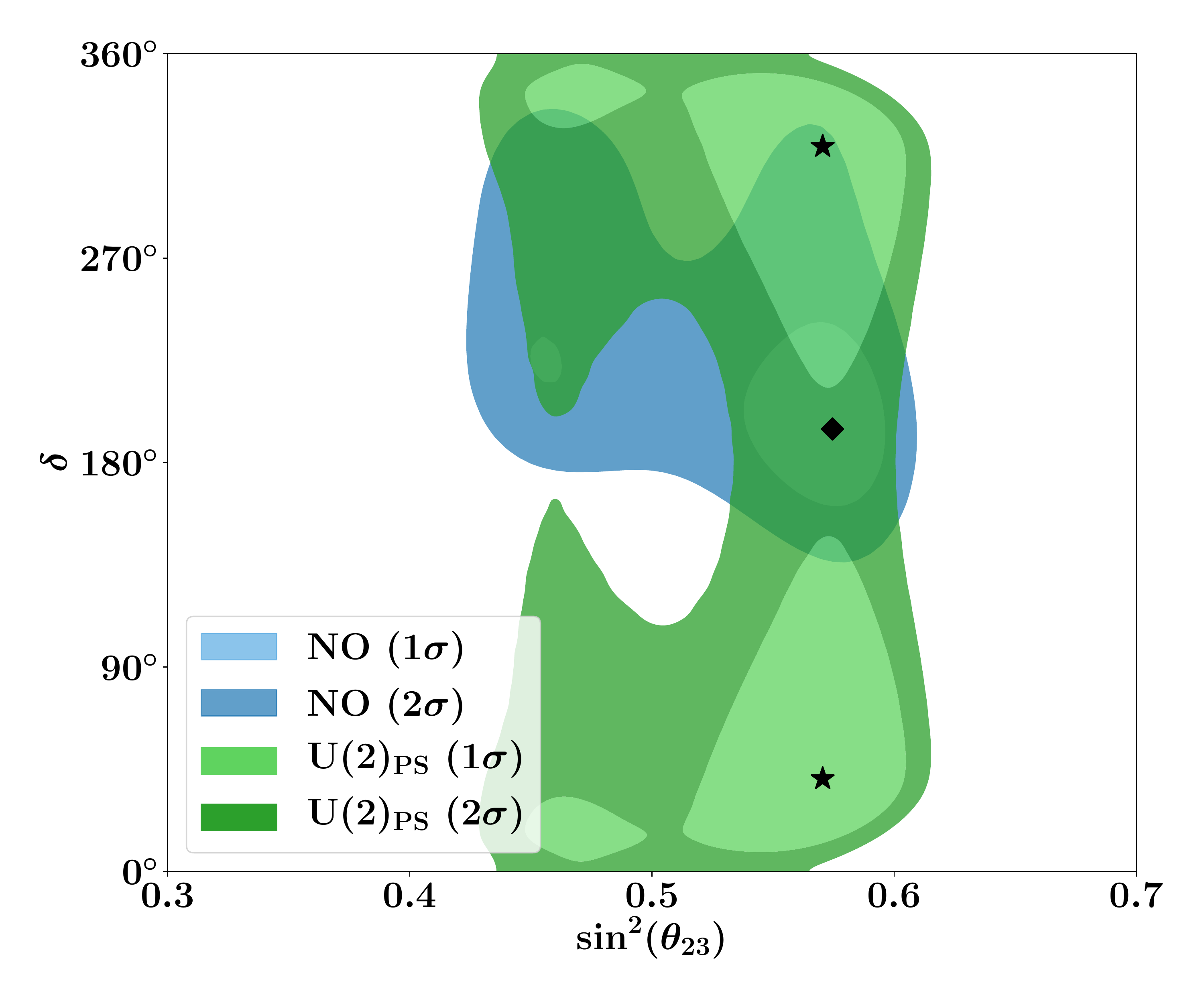}
\includegraphics[scale=0.17]{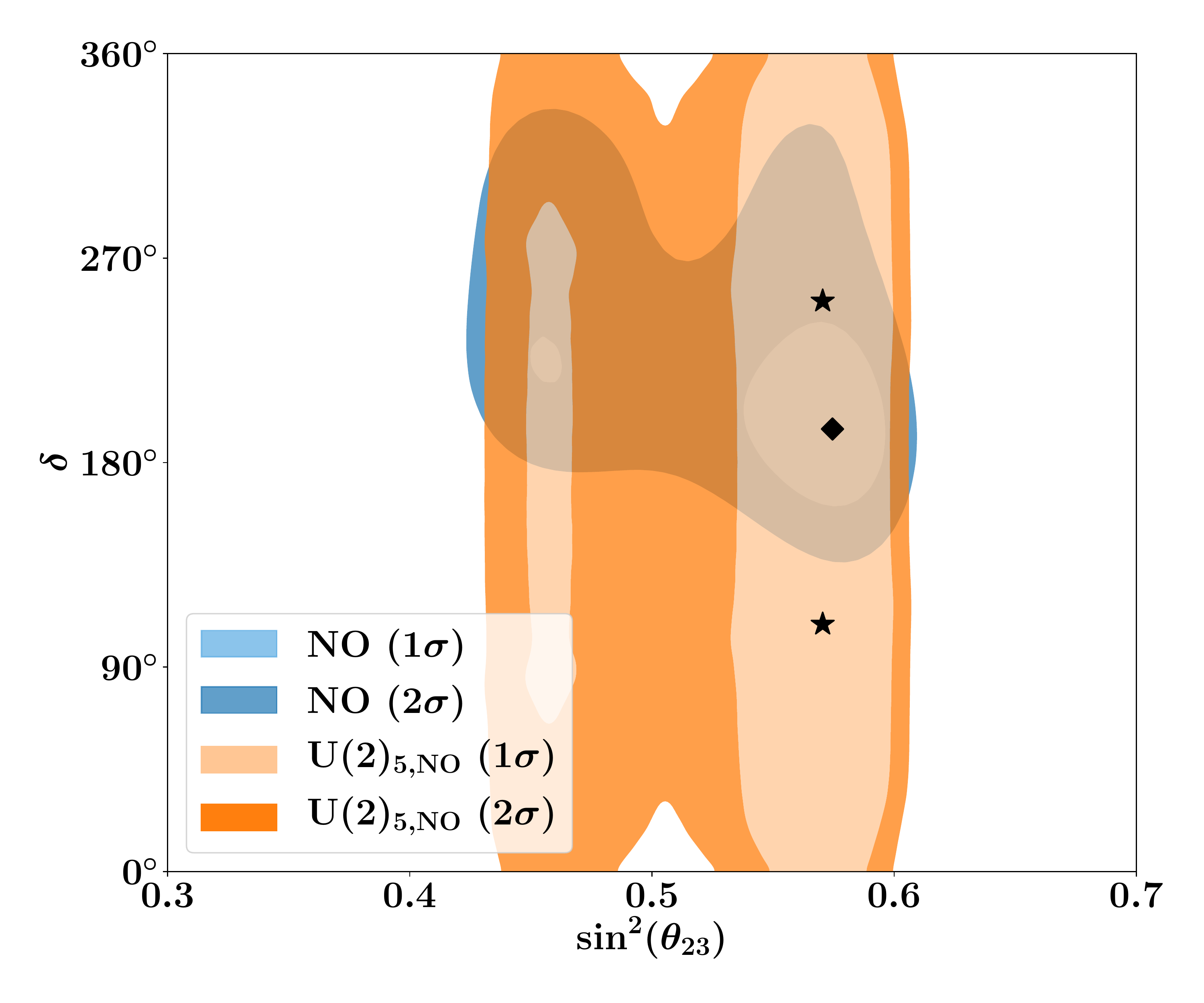}
\includegraphics[scale=0.17]{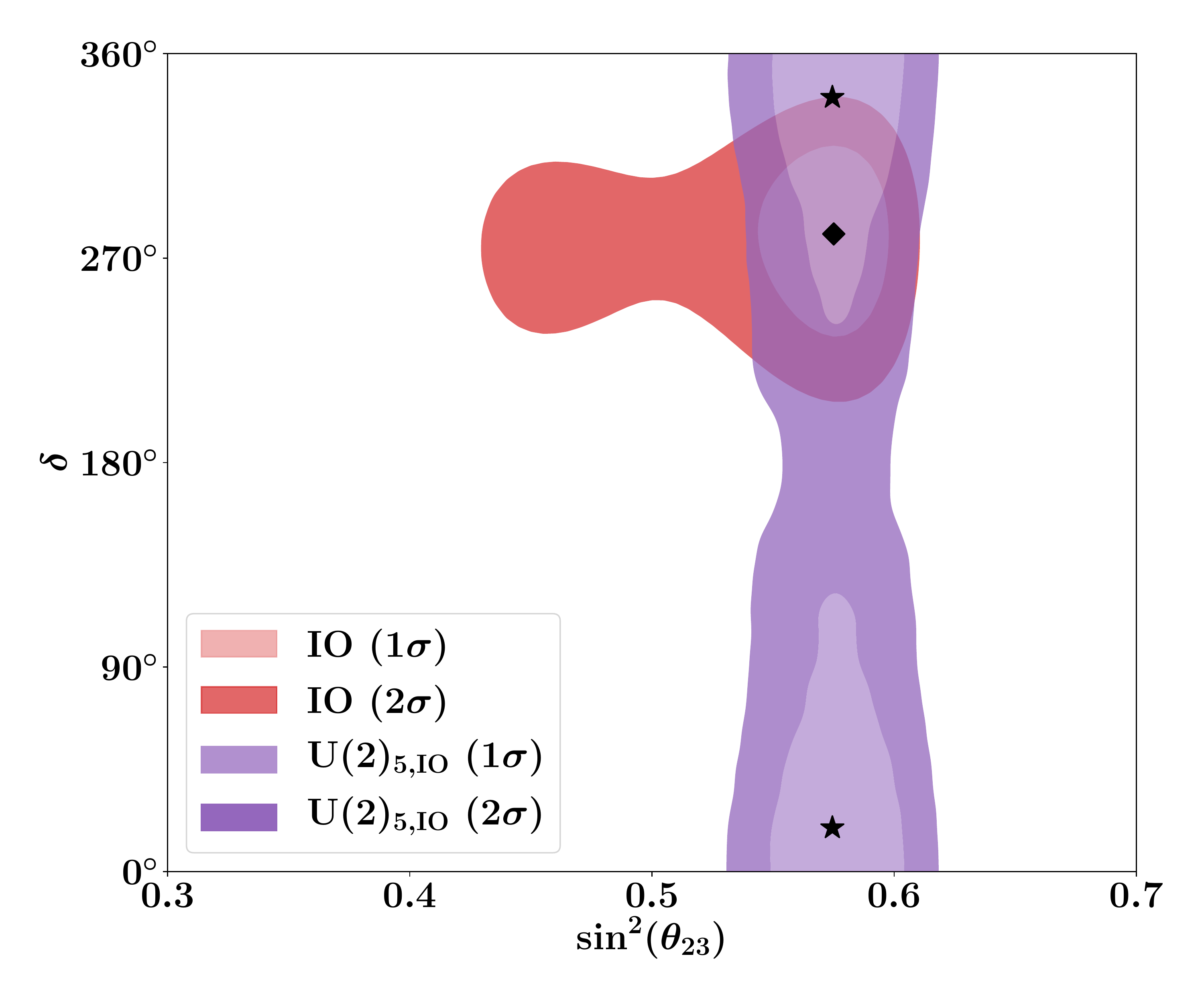}
\caption{\label{CPs23}  Preferred values for the Dirac phase $\delta$ as a function of $\sin^2_{23}$  for the U(2)$_{\rm PS}$   scenario (left panel), the U(2)$_{5, {\rm NO}}$  scenario (central panel) and the U(2)$_{5, {\rm IO}}$ scenario (right panel), where the  black star denotes the most likely value. We overlay these values with the regions preferred by the global fit in Ref.~\cite{Esteban:2020cvm} for NO (blue) and IO (red), where the black diamond is the best-fit point. } 
 \end{figure*}
Our results for the other three scenarios are shown in Fig.~\ref{CPs23}, for U(2)$_{\rm PS}$, U(2)$_{5, {\rm NO}}$ and U(2)$_{5, {\rm IO}}$, respectively. These plots show that in all scenarios the prediction for $\delta$ is not very sharp, as a result of the additional free parameters including new phases. This however depends to some extent on the precise value of $s_{23}^2$, as for special values certain values of $\delta$ can be disfavored. Still much more information is needed in order to draw robust conclusions on these scenarios using their predictions for $\delta$.

%%%%%%%%%%%%%%%%%%%%%% 
\section{Summary and Outlook} 
To summarize, we have studied the predictions  of a U(2) flavor symmetry for neutrino observables. This symmetry predicts a particular structure of mass matrices, which works very well in the quark sector, allowing to account for all hierarchies in quark masses and mixings in terms of just two small parameters. In this article we have assumed the same structure in the neutrino sector and charged lepton sector, which allows to study the resulting predictions for a broad class of U(2) models. While for the neutrino sector the U(2) structure reduces to the well-studied case of the so-called $A_2$ texture zeros, the remaining freedom in the charged lepton sector  can be parametrized by two rotation angles. We have studied the two limits where at least one of these angles is small (which can be motivated in unified scenarios where these rotations are related to small CKM angles), dubbed the ${\rm U(2)_{PS}}$ scenario and the ${\rm U(2)_{5}}$ scenario. When both angles are small, one recovers the predictions of $A_2$ textures, which we have re-analyzed here as the DCL scenario (diagonal charged leptons). This scenario serves as a useful reference point, because its predictions for the overall neutrino mass scale set the ballpark for the other two scenarios. 

Our main results are summarized in Table~\ref{table} and Figs.~\ref{loglog} and \ref{linlin}, which show the viable parameter space for the three scenarios. We find that:
\begin{itemize}
\item IO is viable only for ${\rm U(2)_{5}}$, but the entire range is excluded by cosmology. If IO will be established by upcoming oscillation data, all three scenarios will be excluded.  
\item NO is viable for every scenario, and the resulting range for $\sum  m_i$ falls in a quite narrow window between 60 and 75 meV (at 2$\sigma$), close to the central DCL prediction of 65 meV.  
\item The prediction for $m_{\beta \beta}$ is below 4 meV (at 2$\sigma$), and thus presumably beyond the reach of future $0\nu\beta\beta$ decay experiments. Thus any measurement in the near future will exclude all three scenarios. 
\end{itemize}
While in the DCL scenario there is a sharp prediction for the Dirac phase $\delta$ (cf. Fig.~\ref{CPs23DCL}), in the other two scenarios the presence of extra free parameters smears out this prediction almost  entirely (cf. Fig.~\ref{CPs23}), however this depends to some extent on the true value of $s_{23}$. 

A possible avenue for future studies of these scenarios is the correlation of the free parameters in the charged lepton sector with new observables, for example lepton flavor-violating decays involving a final state (QCD) axion~\cite{Calibbi:2020jvd}, which arises from the breaking of the U(1) factor (and can account both for the strong CP Problem and Dark Matter). If not ruled out by near-future evidence for IO or $0\nu\beta\beta$ decay,  much more precision will be required in order to test these models with neutrino observables. 
%%%%%%%%%%%%%%%%%%%%%% 
\section*{Acknowledgments}
We thank Thomas Schwetz for useful discussions. This work is partially supported  by project C3b of the DFG-funded Collaborative Research Center TRR 257, ``Particle Physics Phenomenology after the Higgs Discovery". ML acknowledges the support by the Doctoral School ``Karlsruhe School of Elementary and Astroparticle Physics: Science and Technology". JLP acknowledge the support from Generalitat Valenciana through the “plan GenT” program (CIDEGENT/2018/019) and from the Spanish MINECO under Grant FPA2017-85985-P.

\appendix

\section{$\mathbf{U(2)_{PS}}$ Relations}
\label{appU2PS}
The charged lepton sector of the $\rm{U(2)_{PS}}$ scenario is parametrized by the angle  $s_{23}^{Re}$ (with boundary $c_{23}^{Re} \gtrsim m_e/m_\mu$) and one free effective phase $\beta$. In terms of these parameters the relations between neutrino sector angles $\theta^\nu_{ij}$ and the physical angles $\theta_{ij}$ are
 \begin{align}
s_{23}^\nu  & = \frac{c_{12}^{Le} c_{13} }{c_{13}^\nu} \left| s_{23}- t_{12}^{Le} t_{13}  e^{i \beta }   \right|  \, , \nonumber \\
s_{12}^\nu   & = \frac{  s_{12}^{Le} s_{12}}{c_{13}^\nu} \left| \frac{c_{13} }{t_{12}^{Le} }   e^{i \beta }   +  \frac{c_{23} }{t_{12}} e^{- i \delta} - s_{23}    s_{13}  \right| \, , \nonumber \\
 s_{13}^\nu   & = c_{12}^{Le} c_{13}   \left| t_{13}  e^{i \beta}  + t_{12}^{Le} s_{23}  \right| \, ,  
 \label{A1}
 \end{align}
 with $s_{12}^{Le} =  \sqrt{c_{23}^{Re}} \sqrt{m_e/ m_\mu}$, while the neutrino phases $\delta^\nu, \alpha^\nu_i$ are related to the physical phases $\delta$ and $\alpha_i$ as
 \begin{align}
   \delta^\nu  &=    \delta + \gamma_1 + \gamma_2  \, ,   \nonumber \\
  \alpha_1^\nu & = \alpha_{1}  - \gamma_1 \, , \nonumber \\
   \alpha_2^\nu  &   = \alpha_{2} - \gamma_1 + \gamma_3 \, , 
   \label{A2}
\end{align}
with the shorthands
\begin{align}
\gamma_1 & =  \arg \frac{ c_{13}/t_{12}^{Le}  e^{i \beta} - s_{23}  s_{13}   + c_{23}/t_{12}   e^{- i \delta}}{s_{23}  e^{ i \beta }  - t_{12}^{Le} t_{13} } \, , \nonumber \\
\gamma_2 & = \arg \frac{c_{13}/t_{12}^{Le}  e^{i \beta } -  s_{23}s_{13}  -  c_{23} t_{12} e^{-i \delta} }{ t_{13}  e^{i \beta} + t_{12}^{Le}  s_{23}} \, , \nonumber \\
\gamma_3 & = \arg \frac{ c_{13}/t_{12}^{Le}  e^{i  \beta   } -   s_{23}  s_{13}  + c_{23}/t_{12} e^{-i \delta}}{c_{13}/t_{12}^{Le} e^{i \beta }  -   s_{23} s_{13}  -  c_{23}  t_{12} e^{-i \delta}}\, .
\label{A3}
\end{align}

\section{$\mathbf{U(2)_{5}}$ Relations}
\label{appU25}
The charged lepton sector in the $\rm{U(2)_{5}}$ scenario is parametrized by the free angle $s_{23}^{Le}$ (with boundary $c_{23}^{Le} \gtrsim m_e/m_\mu$) and two free phases $\beta_{1}, \beta_{2}$. In terms of these parameters the relations between neutrino angles $\theta^\nu_{ij}$ and the physical parameters $\theta_{ij}$ and $\delta$ are:
\begin{align}
s_{23}^\nu  & = \frac{c_{23}^{Le} c_{12}^{Le} c_{13} }{c_{13}^\nu} \left|  s_{23}  - t_{12}^{Le}  t_{13}  e^{i {\beta}_{1} }  +  \frac{c_{23} t_{23}^{Le}}{c_{12}^{Le}}  e^{- i {\beta}_{2} }  \right|  \, , \nonumber \\
s_{12}^\nu   & = \frac{s_{12}^{Le} s_{12} }{c_{13}^\nu} \left|   \frac{c_{13}}{ t_{12}^{Le}} e^{i {\beta}_{1} }    +   \frac{ c_{23}}{t_{12}}  e^{- i \delta}  -   s_{13} s_{23}    \right| \, , \nonumber \\
s_{13}^\nu  & = c_{12}^{Le}  c_{13}   \left| t_{13} e^{i {\beta}_{1}}  + t_{12}^{Le} s_{23}  \right| \, ,  
\label{B1}
\end{align}
with $s^{Le}_{12} = 1/\sqrt{c_{23}^{Le}}  \sqrt{m_e/m_\mu}$, while the neutrino phases $\delta^\nu, \alpha^\nu_i$  are related to physical phases $\delta, \alpha_i$ as
\begin{align}
   \delta^\nu  & =    \delta + {\gamma}_1 +{\gamma}_2  \, ,  \nonumber  \\
     \alpha_1^\nu  & = \alpha_{1}  - {\gamma}_1 \, , \nonumber  \\
      \alpha_2^\nu     & = \alpha_{2} - {\gamma}_1 + {\gamma}_3 \, ,
      \label{B2}
\end{align}
with the shorthands
\begin{align}
{\gamma}_1 & = \arg  \frac{ s_{13} s_{23}  - c_{13}/ t_{12}^{Le} e^{i {\beta}_{1}} - c_{23}/t_{12}  e^{- i \delta} }{t_{13} s_{12}^{Le} +  c_{23}/t_{23}^{Le} e^{i ({\beta}_{1} + {\beta}_{2})} -  s_{23}  c_{12}^{Le} e^{i {\beta}_{1}}} \nonumber \\
& + \arg \left[ t_{13} s_{12}^{Le} e^{i ( {\beta}_{1}+  {\beta}_{2})} -  s_{23} c_{12}^{Le} e^{i {\beta}_{2}} -  c_{23} t_{23}^{Le} \right] \, ,  \nonumber \\
{\gamma}_2 & = \arg  \frac{ -  s_{13} s_{23}  + c_{13}/t_{12}^{Le} e^{i {\beta}_{1}} -  c_{23}/t_{12} e^{- i \delta} }{  s_{23} s_{12}^{Le} + t_{13} c_{12}^{Le} e^{i {\beta}_{1}}} \, , \nonumber \\
{\gamma}_3 & = \arg  \frac{  s_{13} s_{23} - c_{13}/t_{12}^{Le} e^{i {\beta}_{1}} - c_{23}/t_{12} e^{- i \delta} }{ s_{13} s_{23}  - c_{13}/t_{12}^{Le} e^{i {\beta}_{1} } + t_{12} c_{23}  e^{-i \delta }} \, . 
\label{B3}
\end{align}

\bibliographystyle{apsrev4-1.bst}
%%%%%%%  \bibliographystyle{utphys.bst}
%\bibliography{bibliography}
\bibliography{refs}

\end{document}